\documentclass{PoS}

\title{Vector boson production in association with jets and heavy flavor quarks from CMS}

\ShortTitle{V+jets measurements from CMS}

\author{\speaker{Fengwangdong Zhang}\\
	      On behalf of the CMS Collaboration \\
        IIHE - Universit\'e Libre de Bruxelles (BE), Peking University (CN)\\
        E-mail: \email{fengwangdong.zhang@cern.ch}}

\abstract{This document is dedicated to recent results on the measurements of a 
vector boson production associated with jets (V+jets) using CMS detector experiment, with a 	
central proton-proton (pp) collision energy of respectively 8 TeV (RunI) and 13 TeV (RunII).
The vector boson can be a W boson, Z boson, or photon($\gamma$). The jets are of any flavor, as well 
as the ones containing heavy flavor quarks, such as bottom(b) quarks. 
The inclusive cross sections of V+jets processes and differential cross sections as a function of 
various kinematic observables are presented and compared to several theoretical predictions. In general, 
the comparison between theoretical calculations and measurement are crucial for improving modeling and precision 
on QCD dynamics.}

\FullConference{XXIV International Workshop on Deep Inelastic Scattering and Related Subjects\\
		11--15 April 2016\\
		DESY, Hamburg, Germany}

\begin{document}

\section{Motivation on V+jets physics}
The production of a vector boson associated with hadronic jets is a crucial process at the LHC, 
which provides a reference both for physics measurements and detector response. 
Since it involves Quantum Chromodynamics (QCD) interactions, it can test different aspects 
of QCD computations. In parallel, a vector boson, which can be Z boson, W boson, or photon, produced 
through electroweak interaction, provides 
important signature of the event. Therefore, precision measurements of these processes are 
indispensable. Actually, our understanding and modeling of QCD interactions can affect the accuracy levels 
when we proceed with theoretical predictions.

Furthermore, W/Z+jets process is a dominant background for such as top quark production, Higgs physics precision 
measurements, and searches for new physics, such as searches for Supersymmetry signatures. 
It may improve the purity and sensitivity of these relatively 
rare processes if one can measure and control the W/Z+jets process precisely enough, which can also 
help to constrain the theoretical modeling of data.

We will present V+jets measurements performed by the CMS collaboration \lbrack1\rbrack. 

\section{Measurement of the Z+jets cross section}
The production of Z bosons in association with jets has a quite clean signature with low background 
contamination, and a high reconstruction efficiency of Z bosons, selected here with a decay in two oppositely 
charged leptons. Thus it offers ideal conditions for the study of hadronic jet production. By measuring 
the cross section, one can realize the stringent tests of perturbative QCD and of the relevant Monte 
Carlo (MC) simulations. The data for this measurement were taken by the CMS detector with a luminosity 
of 19.6 $\rm fb^{-1}$ in 2012 and 2.5 $\rm fb^{-1}$ in 2015. 

\subsection{Z+jets at 8TeV \lbrack2\rbrack}
The measurement of differential Z+jets cross sections at 8 TeV are compared with several theoretical predictions. 
One of the theoretical computations is a multi-leg leading order (LO) calculation with 0-4 final 
partons in the matrix element. The dedicated generator is MADGRAPH5 + PYTHIA6 ($\rm k_{T}$-MLM merging, CTEQ6L1 PDF set) \lbrack3,4\rbrack.
The parton showers are matched with matrix elements. The non-perturbative effect, including 
hadronization and multiple parton interaction (MPI), are also implemented in these generators. 
The second prediction is multi-leg with 0-2 partons at next leading order (NLO) level, and with 3-4 partons 
at LO level. The corresponding generator is SHERPA2 + BLACKHAT (MEPS@NLO merging, CT10PDF) \lbrack5\rbrack.
As before, the parton shower and non-perturbative effects are implemented in this generator.
For next leading order level, the SHERPA2 is used for the Born calculation and real emission corrections, 
while BLACKHAT is used for the virtual loop corrections.

In order to reconstruct a pure sample of Z bosons, two opposite charged same flavor leptons (either muons or 
electrons) are required. In addition, one or more reconstructed jets
are selected. The specific selection criteria are:
\begin{itemize}
 \setlength\itemsep{0.001em}
 \item $p_{T}^{l} > $20 GeV (lepton transverse momentum), $|\eta^{l}| < $2.4 (lepton absolute pseudorapidity)
 \item 71$ < M_{ll} < $111 GeV (invariant mass of the lepton pair)
 \item $p_{T}^{j} > $30 GeV (jet transverse momentum), $|\eta^{j}| < $2.4 (jet absolute pseudorapidity)
\end{itemize}

Based on this kinematic selection, one measures the differential cross section as 
a function of different observables. The comparisons between measurements and theoretical predictions 
at 8 TeV are shown in figure 1. One can see the general good agreement, and the discrepancy with LO 
approximation has disappeared with the NLO accuracy. The overshooting for high jet multiplicities stems from 
the fact that SHERPA2 relies on parton shower for the modeling of high jet multiplicities.

\begin{figure}
{\includegraphics[width= 0.49\textwidth,height=8.5cm]{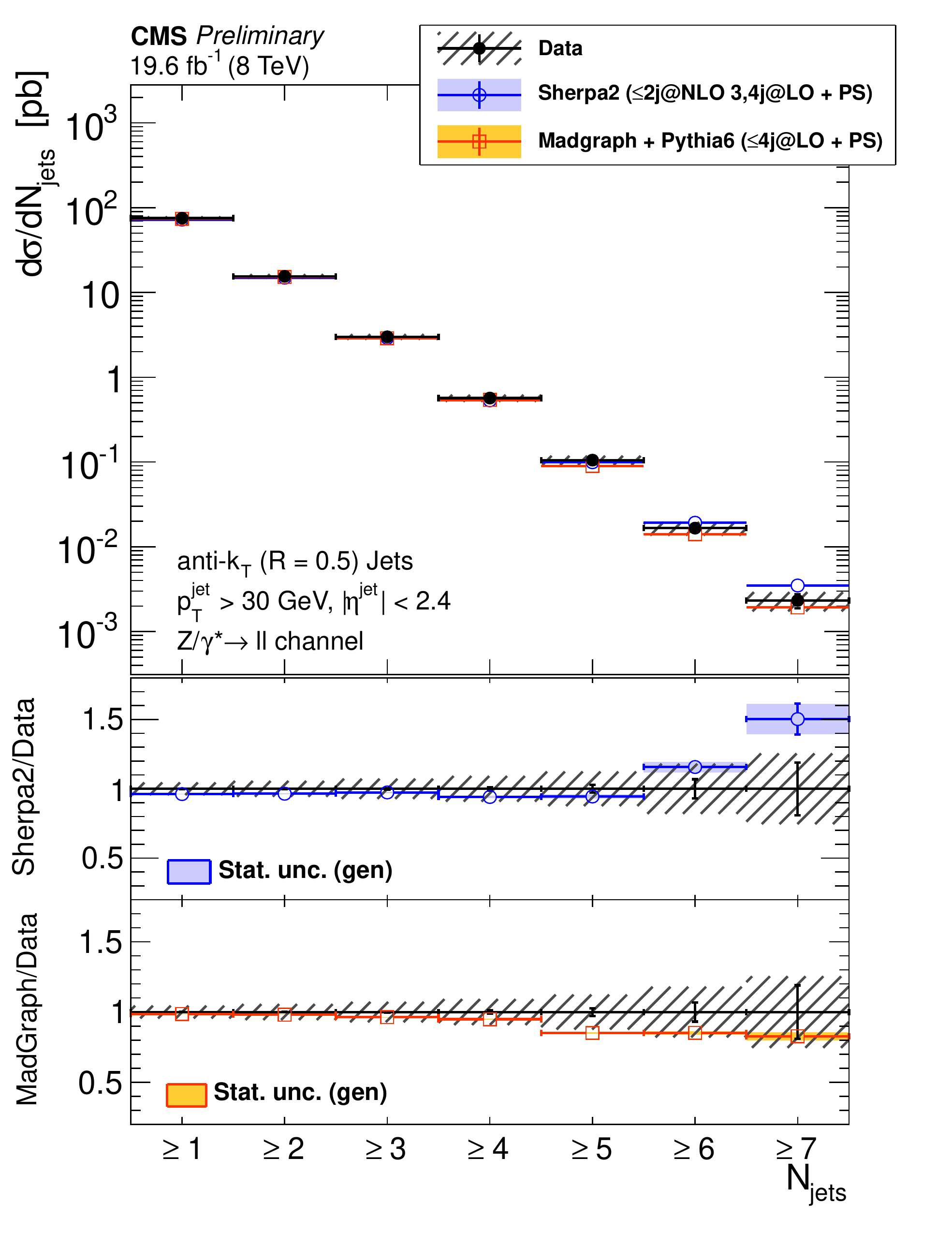}}
{\includegraphics[width= 0.49\textwidth,height=8.5cm]{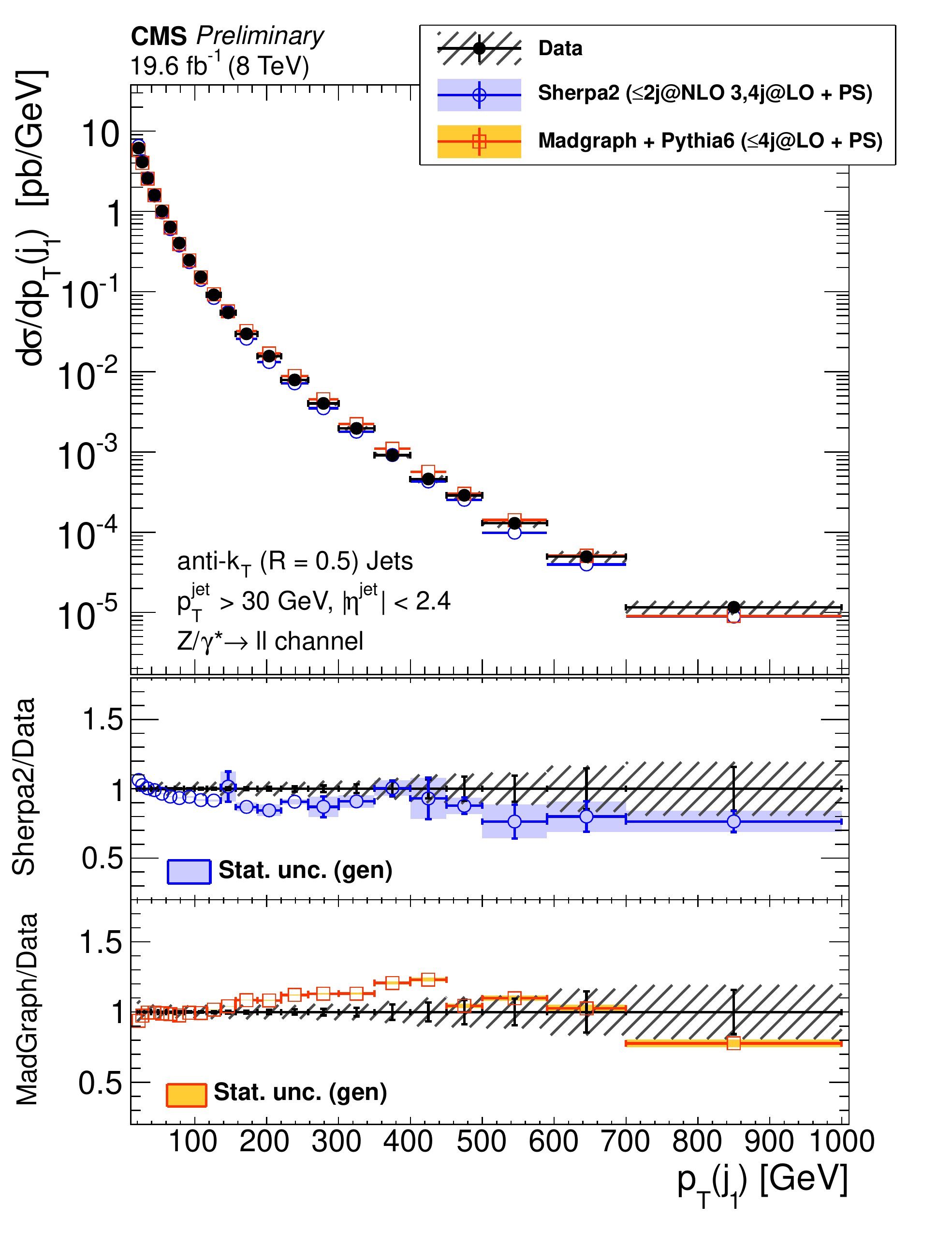}}
\label{fig:zjets8tev}
\caption{The differential cross section of Z+jets production at 8 TeV as a function of inclusive jet multiplicity 
(left), and of leading jet $p_{T}$ (right). The measurement is compared with MADGRAPH(LO) 
and SHERPA2(NLO) estimations. The error bars represent the statistical uncertainties of data 
and theoretical predictions. The color bands are the statistical uncertainties of predictions, and the hashed bands 
are the quadrature of systematic uncertainty and statistical uncertainty.}
\end{figure}

\subsection{Z+jets at 13TeV \lbrack6\rbrack}
The measurements of differential Z+jets cross sections at 13 TeV are implemented with the same kinematic 
selection as the ones at 8 TeV, and also compared with several theoretical predictions. The first theoretical computation is a NLO multi-leg matrix element calculation up to two jets, 
and LO accuracy for 3-4 jets. They are matched with parton shower by FxFx merging scheme. 
The hadronization and MPI effects are included in this generator. 
This scenario is realized by MADGRAPH5\_aMC@NLO \lbrack7\rbrack interfaced with PYTHIA8 \lbrack8\rbrack. 
The second calculation used is fixed next-to-next-leading-order (NNLO) computation for 
Z plus one jet. Since the cross section is calculated at parton level, the 
corrections for hadronization and MPI are estimated independently with MADGRAPH5\_aMC@NLO and PYTHIA8, 
and applied to the NNLO predictions. 
The comparison between data and the predictions is shown in figure 2. 
In general, one can see good agreement at both NLO and NNLO accuracy levels. 
The $p_{T}$, $\eta$, and $H_{T}$(scalar sum of jets) for inclusive jet 
multiplicity up to three have also been measured.

\begin{figure}
{\includegraphics[width= 0.49\textwidth,height=8.5cm]{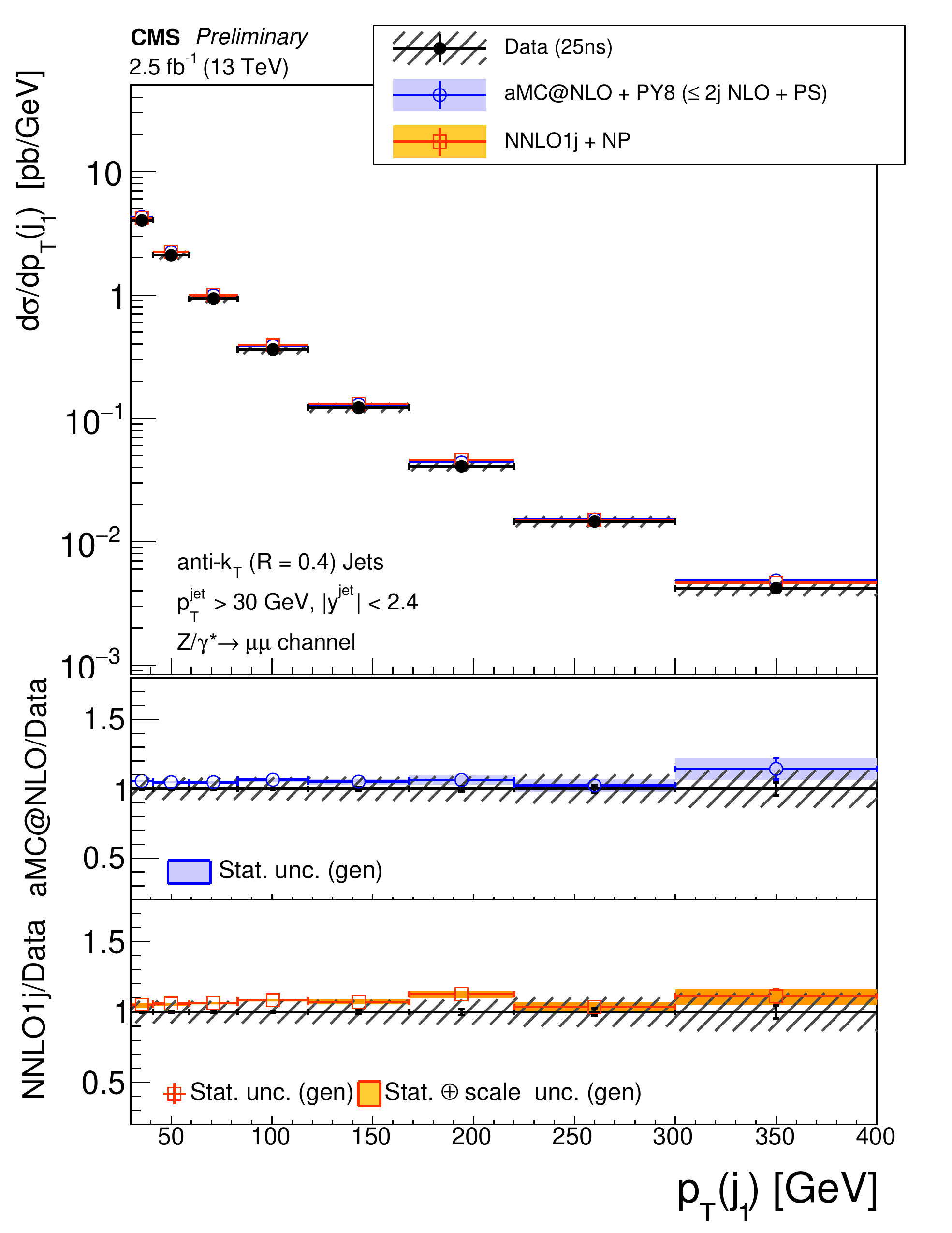}}
{\includegraphics[width= 0.49\textwidth,height=8.5cm]{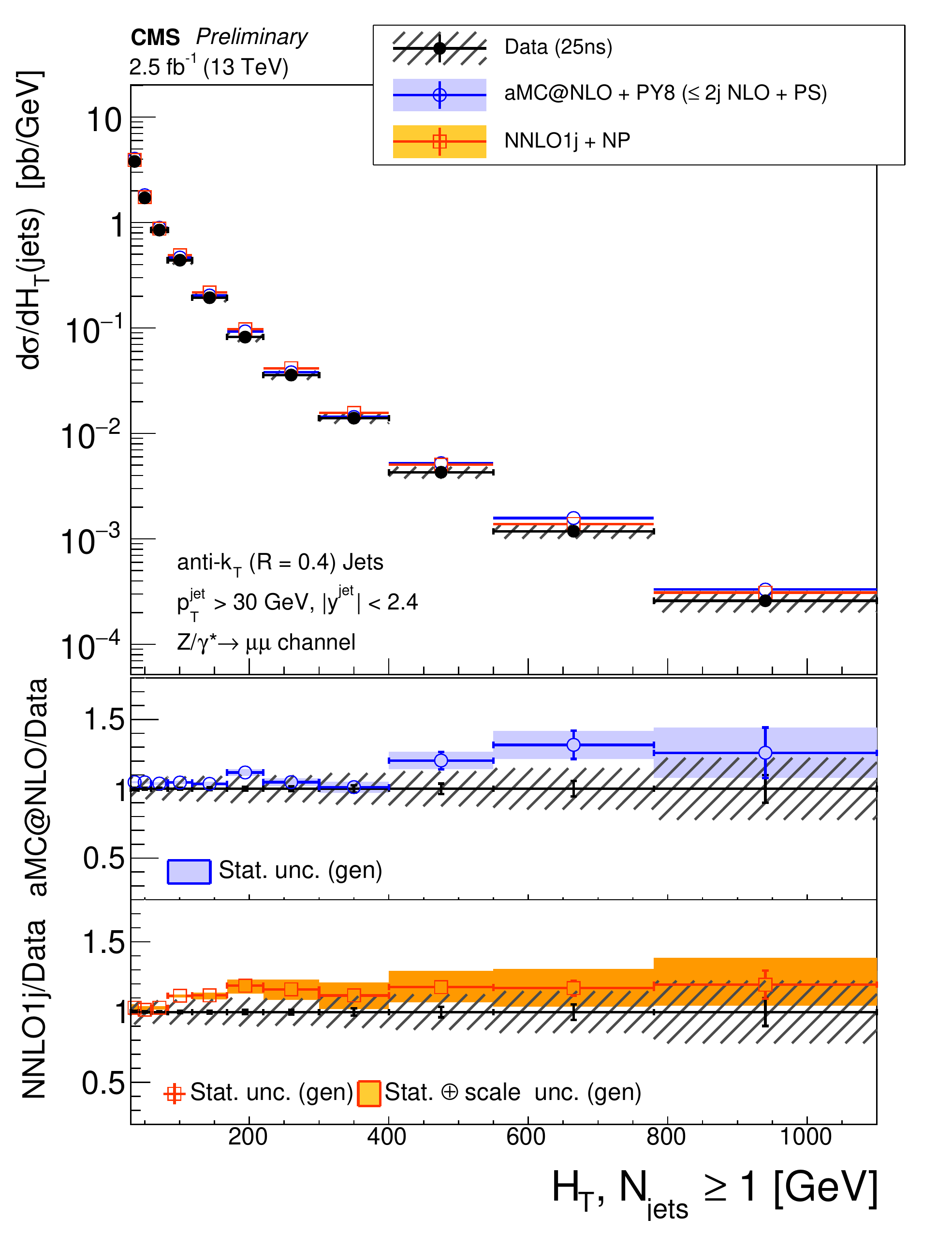}}
\caption{The differential cross section of Z+jets production at 13 TeV as a function of leading jet $p_{T}$ (left) 
and scalar sum ($H_{T}$) of jets $p_{T}$ (right). The error bars represent the 
statistical uncertainties of data and predictions. The blue band is the 
statistical uncertainty of aMC@NLO, while the orange band corresponds to the 
quadrature of the statistical uncertainty of non-perturbative correction and 
theoretical scale uncertainty of NNLO. 
The hashed bands are the quadrature of systematic uncertainty and statistical uncertainty.}
\label{fig:zjets13tev}
\end{figure}

\section{Measurement of $\gamma$+jets cross section at 8 TeV \lbrack9\rbrack}
Photon production associated with jets has similar QCD topology to Z+jets, 
therefore it is also a good probe on perturbative QCD calculations. By computing 
the ratio between Z+jets and $\gamma$+jets cross sections, one can measure hadronic
production with reduced systematical uncertainty. Compared to Z+jets, 
the $\gamma$+jets process has larger cross section, because the electromagnetic coupling is 
much stronger than the weak coupling. Furthermore, Z+jets and $\gamma$+jets processes 
are easy to be distinguished due to the different masses of Z bosons and photons. 

The measurement is compared with LO and NLO calculations. The specific 
generators are MADGRAPH5 + PYTHIA6 and BLACKHAT + SHERPA (MSTW2008 NLO PDF) \lbrack10\rbrack. The former is based 
on tree level matrix element and it is interfaced with parton 
shower. The latter is fixed order computation, implementing virtual loop 
correction by means of BLACKHAT, and using Born level calculation and real emission corrections 
implemented by SHERPA. 

For the comparison, fiducial selections are applied:
\begin{itemize}
 \setlength\itemsep{0.001em}
 \item $p_{T}^{\gamma} > $30 GeV (photon transverse momentum), $|y^{\gamma}| < $1.4 (photon absolute rapidity)
 \item $p_{T}^{l} > $20 GeV (lepton transverse momentum), $|\eta^{l}| < $2.4 (lepton absolute pseudorapidity)
 \item $p_{T}^{j} > $30 GeV (jet transverse momentum), $|\eta^{j}| < $2.4 (jet absolute pseudorapidity)
\end{itemize}
The comparison between measurements and predictions is shown in figure 3. For the 
differential cross section, the NLO prediction apparently overcomes the undershooting 
of the LO computation in the low $p_{T}^{\gamma}$ region. 

\begin{figure}
{\includegraphics[width= 0.49\textwidth, height=5.5cm]{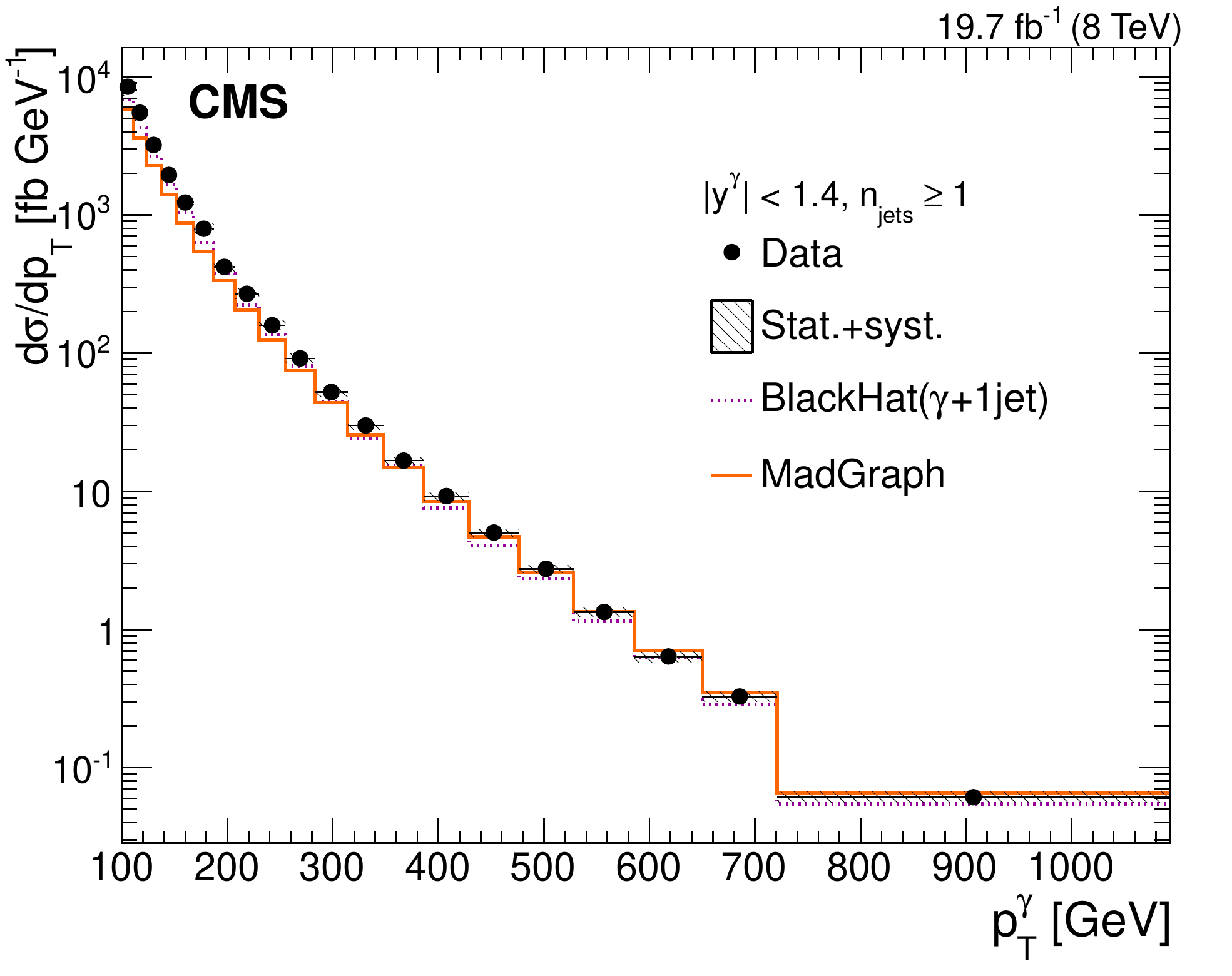}}
{\includegraphics[width= 0.49\textwidth, height=6.1cm]{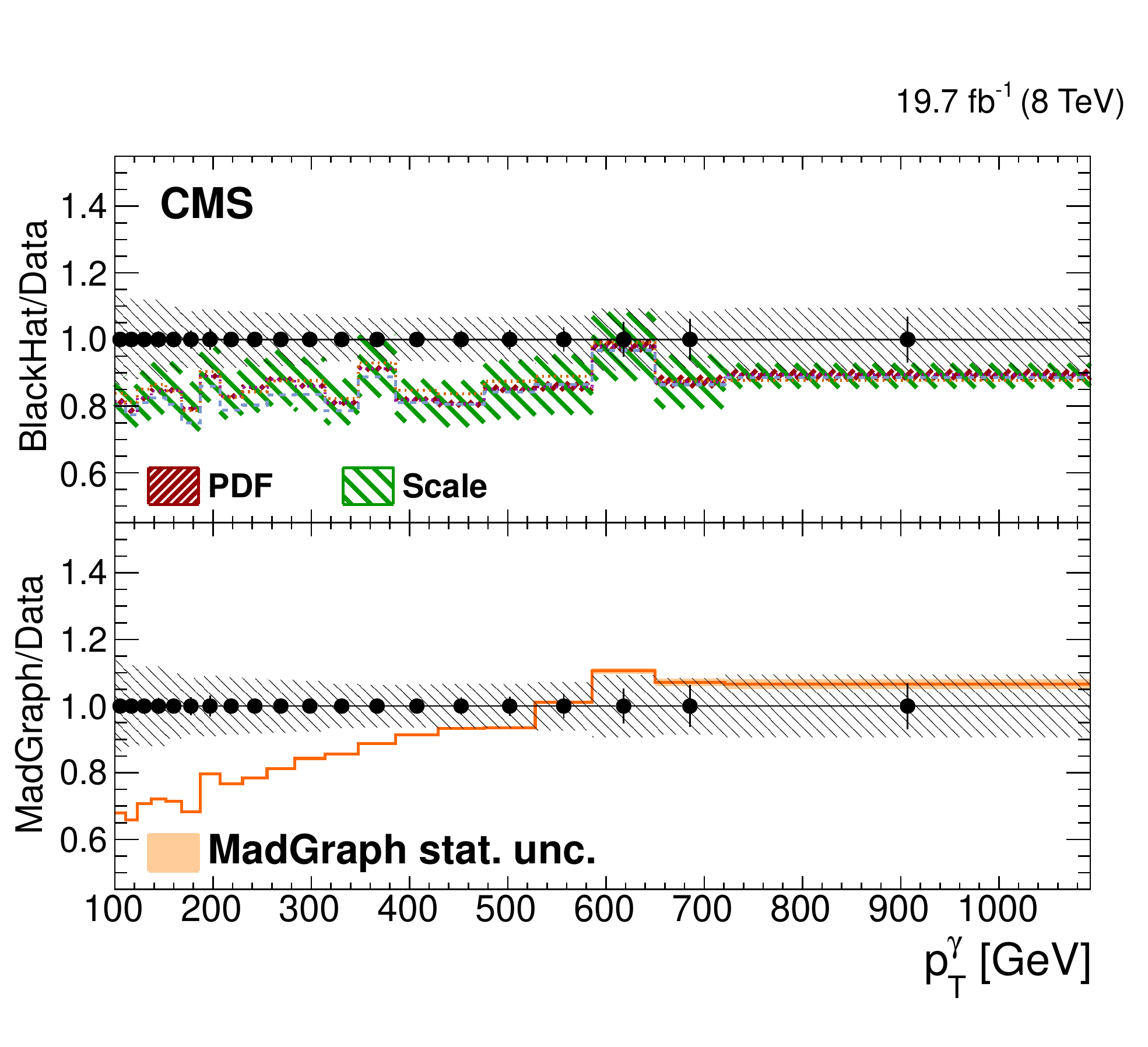}}
\caption{The differential cross section of $\gamma$+jets production at 8 TeV as a function of $p_{T}^{\gamma}$ (left) and 
the ratio between theoretical predictions and measurement (right). The measurement is compared with 
MADGRAPH(LO) and BLACKHAT(NLO) computations. The error bars are the statistical uncertainty of data, and the hashed black bands 
correspond to the quadrature of statistical uncertainty and systematical uncertainty. 
The hashed light green bands are the scale variation uncertainty of BLACKHAT, while the hashed 
deep green bands stand for the parton density function(PDF) uncertainty.}
\label{fig:photonjet}
\end{figure}

\section{Measurement of W/Z + b-quark jets production at 8 TeV \lbrack11,12\rbrack}
W/Z boson production in association with b-quark jets is an important background of Higgs boson 
measurements, such as Higgs production with a vector boson, where Higgs decays 
to b$\rm\bar{b}$. There are two different 
schemes, which are respectively called the four flavor number scheme (4FS/4F) and five 
flavor number scheme (5FS/5F). The former generates massive b-quarks by gluon 
splitting, while the latter includes massless b-quarks in the initial state 
of proton. The modeling of b-quark production leads to larger theoretical uncertainties 
than the modeling of light flavor quarks.

For the W+b$\rm\bar{b}$ inclusive cross section measurement, 
the MCFM (NLO) \lbrack13\rbrack and MADGRAPH (LO) generators are used for comparison. 
The LO prediction is interfaced with PYTHIA6 respectively in 4FS 
and 5FS. Meanwhile, PYTHIA8 with 4FS is also used for comparison. Since the MCFM 
computation is based on parton level, hadronization and MPI corrections are derived and applied to the 
MCFM predictions. 
Note that the b$\rm\bar{b}$ production resulting from double parton scattering 
is not considered in MCFM and 4FS, thus there is an additional correction factor for that. 
The kinematic selections are:
\begin{itemize}
 \setlength\itemsep{0.001em}
 \item $p_{T}^{l} >$ 30 GeV (lepton transverse momentum), $|\eta^{l}| < $2.1 (lepton absolute pseudorapidity)
 \item $p_{T}^{b} >$ 25 GeV (b-quark jet transverse momentum), $|\eta^{b}| < $2.4 (b-quark jet absolute pseudorapidity)
\end{itemize}
According to figure 4, all the theoretical predictions are consistent with measurement 
within one standard deviation.

Regarding Z+b(b), differential cross sections have been measured, which are compared 
with MADGRAPH5(LO)+PYTHIA6 respectively in 4F and 5F, and with POWHEG(NLO)+PYTHIA6 \lbrack14\rbrack. 
The basic kinematic cuts are the same as Z+jets in Sec.2, except the choice of the 
jet flavor. From the comparison in figure 4, one can see a discrepancy in shape when compared 
with 4F, with a small overshooting for the soft leading b-quark jet. On the 
other hand, the NLO calculation is in good agreement with data.

\begin{figure}
\begin{center}
{\includegraphics[width= 0.49\textwidth, height=5.5cm]{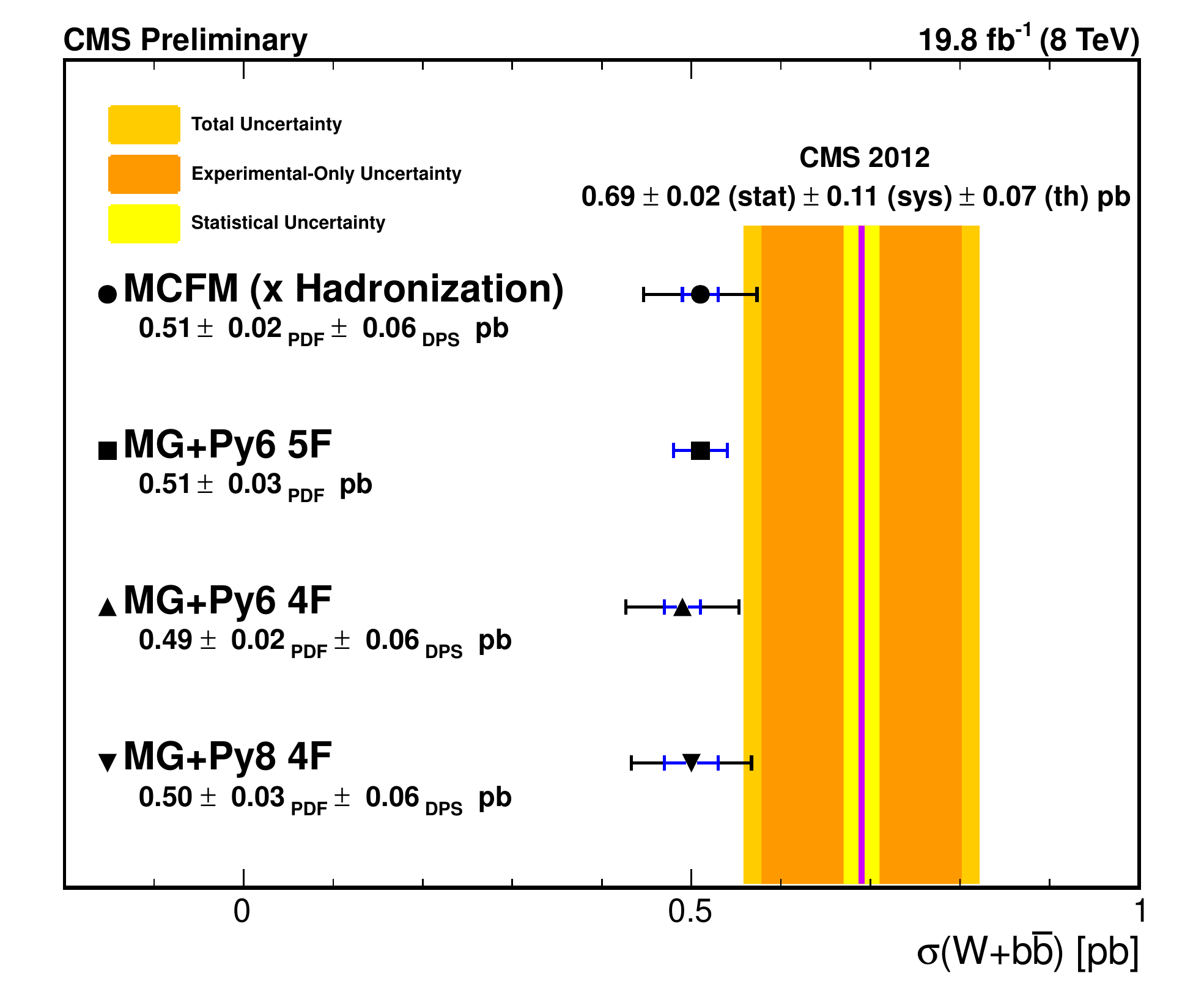}}
{\includegraphics[width= 0.49\textwidth, height=5.5cm]{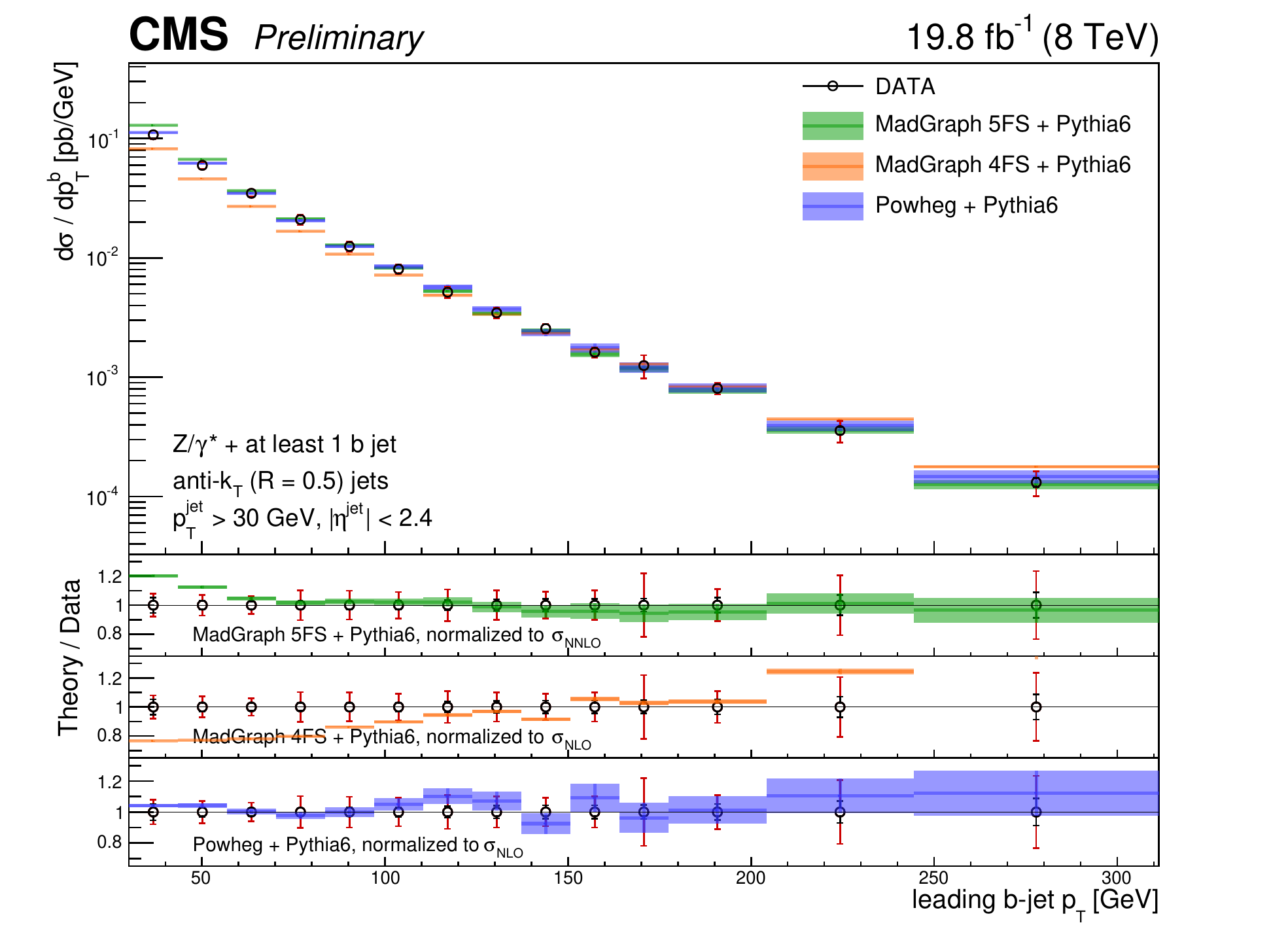}}
\caption{\newline{\bf Left:} Comparison between the measured W+b$\rm\bar{b}$ inclusive cross section at 8 TeV 
and various QCD theoretical predictions. The blue error bars on the predictions represent the 
uncertainty in the given sample associated with PDF choice and the black bars represent 
the total uncertainty. In the case of the MADGRAPH+PYTHIA6(5F) sample, the effects of double 
parton scattering (DPS) 
are already included in the generated sample, so the extra DPS factor is not needed.
\newline{\bf Right:} Measured differential Z(1b) cross section as 
a function of the leading b-quark jet $p_{T}$, compared with
the MADGRAPH(5FS), MADGRAPH(4FS) and POWHEG theoretical predictions (shaded bands). 
For each data point the statistical and the total (statistical plus systematic) uncertainties 
are represented by the double error bar. The width of the shaded bands represents the statistical 
error on the theoretical predictions.}
\label{fig:meteffi}
\end{center}
\end{figure}

\section{Summary}
The standard model V+jets cross sections, at 8 TeV or 13 TeV, 
have been measured by CMS collaboration and compared with 
different theoretical predictions. The measurements are becoming increasingly more precise, and 
the comparison with theoretical predictions is improved by the availability of higher order calculations.

\end{document}